\documentclass{iopart}
\usepackage{iopams} 
\usepackage{epsfig}  
\usepackage{graphics}  
\begin{document}

\title{Matter waves from quantum sources in a force field}

\author{T. Kramer, C. Bracher and M. Kleber}

\address{Physik-Department T30,
Technische Universit\"at M\"unchen,
James-Franck-Stra{\ss}e,
85747 Garching, Germany}

\ead{tkramer@ph.tum.de}

\begin{abstract}
Localized scattering phenomena may result in the formation of stationary matter waves originating from a compact region in physical space. Mathematically, such waves are advantageously expressed in terms of quantum sources that are introduced into the Schr\"odinger equation. The source formalism yields direct access to the scattering wave function, particle distribution, and total current. As an example, we study emission from three-dimensional Gaussian sources into a homogeneous force field. This model describes the behaviour of an atom laser supplied by an ideal Bose-Einstein condensate under the influence of gravity. We predict a strong dependence of the beam profile on the condensate size and the presence of interference phenomena recently observed in photodetachment experiments.
\end{abstract}

% 03.75.-b ... Matter waves
% 03.65.Nk ... Scattering theory
% 32.80.Gc ... Photodetachment of negative atomic ions

\submitto{\JPA}

%\maketitle

\section{Introduction}

Sources of particles are known to be an indispensable element in all scattering experiments. Located far away from the scattering region, sources will give rise to boundary conditions on the scattering wave function in the form of incoming plane waves. However, sources that are close to or within the relevant interaction region should be introduced directly into the Schr\"odinger equation \cite{Rodberg67}. More generally, the concept of a particle source arises naturally if a complex quantum process can be decomposed into several steps, where the preparation of the state under consideration is separated from its further evolution in an external field. For example, in intense-field laser-atom physics, the laser interacts with a pool of atoms in a complicated nonlinear way. As a result of that process a coherent source of above-threshold electrons is generated that will emit electrons in the form of an outgoing spherical wave with a given energy in the continuum \cite{Muller90,Becker94,Lohr97}. 

In this article we discuss in more detail a stationary quantum source that allows for a fully analytical solution. Our novel example predicts how the properties of the resulting matter waves are controlled by the size of the underlying source. The model we are presenting is surprisingly simple: A weak, harmonic perturbation couples the bound state in an isotropic oscillator potential to a continuum state experiencing a homogeneous force field. In section~\ref{sec:SourceModel}, we show how the ground state wave function acts as a stationary quantum source for the uniformly accelerated scattering wave, and discuss the ensuing integrated current and current distribution. For a three-dimensional Gaussian source embedded into a linear potential environment, an exact solution for the emitted wave and the  associated current is obtained in section~\ref{sec:SourceUniformField}, which allows for a pictorial interpretation in terms of a ``virtual'' point source. In section~\ref{sec:Examples}, this theory serves as a simple model for two phenomena recently investigated experimentally, the formation of a freely falling ``atom laser'' beam from a trapped Bose-Einstein condensate, as well as near-threshold photodetachment microscopy of negative ions in an electric field. Finally, a brief outlook is given in section~\ref{sec:Outlook}.

\section{Quantum sources in an external field}\label{sec:SourceModel}

To motivate the quantum source approach, we will introduce a two state model with time-independent Hamiltonians. One state is a bound state of the Hamiltonian $H_{\rm trap}$, the second state is a scattering state of the Hamiltonian $H_{\rm cont}$ containing an external potential. Both states have an energy difference of $\Delta E=E_{\rm cont}-E_{\rm trap}$ and are weakly coupled by a homogeneous but oscillating interaction potential of strength $\hbar\Omega$:
\begin{eqnarray}
(\rmi\hbar\partial_t-H_{\rm cont})\psi_{\rm cont}(\bi{r},t)
 &=\hbar\Omega\rme^{-\rmi\Delta E t/\hbar}\psi_{\rm trap}(\bi{r},t) \\
(\rmi\hbar\partial_t-H_{\rm trap})\psi_{\rm trap}(\bi{r},t)
 &=\hbar\Omega\rme^{+\rmi\Delta E t/\hbar}\psi_{\rm cont}(\bi{r},t).
\end{eqnarray}
We split off the time dependence of the states
\begin{eqnarray}
\psi_{\rm cont}(\bi{r},t)&=\rme^{-iE_{\rm cont}t/\hbar}\psi_{\rm cont}(\bi{r})\\
\psi_{\rm trap}(\bi{r},t)&=\rme^{-iE_{\rm trap}t/\hbar}\psi_{\rm trap}(\bi{r}),
\end{eqnarray}
to obtain the stationary equations
\begin{eqnarray}
(E_{\rm cont}-H_{\rm cont})\psi_{\rm cont}(\bi{r})
 &=\hbar\Omega\psi_{\rm trap}(\bi{r}) \label{eq:PsiStationaryCont}\\
(E_{\rm trap}-H_{\rm trap})\psi_{\rm trap}(\bi{r})
 &=\hbar\Omega\psi_{\rm cont}(\bi{r}).
\end{eqnarray}
Upon introducing a suitable energy-dependent Green function $G_{\rm cont}(\bi{r},\bi{r}';E)$ for $H_{\rm cont}$ (see section~\ref{sec:GreenAnalytic})
\begin{equation}
(E-H_{\rm cont})G_{\rm cont}(\bi{r},\bi{r}';E)=\delta(\bi{r}-\bi{r'}),
\end{equation}
the formal solution of equation~(\ref{eq:PsiStationaryCont}) is given by
\begin{equation}\label{eq:PsiContGreen}
\psi_{\rm cont}(\bi{r})=\hbar\Omega \int \rmd^3\bi{r}'\,G_{\rm cont}(\bi{r},\bi{r}';E_{\rm cont})\,\psi_{\rm trap}(\bi{r}').
\end{equation}
A similar equation holds for $\psi_{\rm trap}(\bi{r})$.
However, if we assume only a weak interaction in the sense that $\psi_{\rm trap}(\bi{r})$ is not changed appreciably by the interaction, we may as a first approximation replace $\psi_{\rm trap}(\bi{r})$ in equation~(\ref{eq:PsiContGreen}) by the bound eigenstate $\psi_0(\bi{r})$ of $H_{\rm trap}$, which is defined by
\begin{equation}\label{eq:Psi0}
(E_0-H_{\rm trap})\psi_0(\bi{r})=0.
\end{equation}
In this way we have decoupled both equations and we obtain a new Schr\"odinger equation with an inhomogeneous source term $\sigma(\bi{r})=\hbar\Omega\psi_0(\bi{r})$
\begin{equation}\label{eq:SchroedingerSource}
(E_{\rm cont}-H_{\rm cont})\psi_{\rm cont}(\bi{r})=\sigma(\bi{r}).
\end{equation}
From now on we will suppress the label ${\it cont}$, since we are not interested in the propagation of the bound state. Therefore the rewritten equation~(\ref{eq:PsiContGreen})
\begin{equation}\label{eq:Wavefunction}
\psi(\bi{r})=\int\rmd^3\bi{r}'\,G(\bi{r},\bi{r}';E)\,\sigma(\bi{r}').
\end{equation}
serves as the starting point for our development of the theory of quantum sources.

\subsection{Currents from sources}

We associate a probability current density $\bi{j}(\bi{r})$ with the scattering wave $\psi(\bi{r})$ via
\begin{equation}\label{eq:CurrentDensity}
\bi{j}(\bi{r})=
\frac{\hbar}{m}{\rm Im}\{\psi(\bi{r})^{*}\,\nabla \psi(\bi{r})\}-\frac{e \bi{A}(\bi{r})}{m}{|\psi(\bi{r})|}^2,
\end{equation}
where $\bi{A}(\bi{r})$ denotes the vector potential. Together with equation~(\ref{eq:SchroedingerSource}) it is straightforward to derive the equation of continuity for this stationary problem:
\begin{equation}\label{eq:CurrentDivergence}
\nabla\cdot\bi{j}(\bi{r})=-\frac{2}{\hbar}{\rm Im} \{\sigma(\bi{r})^{*}\,\psi(\bi{r})\}.
\end{equation}
Note that the introduction of sources $\sigma(\bi{r})$ in the Schr\"odinger equation~(\ref{eq:SchroedingerSource}) causes the appearance of a source term in the equation of continuity that depends on the wave function $\psi(\bi{r})$. By integration over a surface enclosing the source we obtain a bilinear expression for the total probability current carried by $\psi(\bi{r})$:
\begin{equation}\label{eq:CurrentTotal}
J(E)=-\frac{2}{\hbar}{\rm Im}
\left\{
\int\rmd^3\bi{r}\,\int\rmd^3\bi{r}'\,\sigma(\bi{r})^{*}\,G(\bi{r},\bi{r}';E)\,\sigma(\bi{r}')
\right\}.
\end{equation}
This quantity details the total cross section of the underlying scattering process.

\subsection{The energy-dependent Green function}\label{sec:GreenAnalytic}

For our stationary problem, the mathematical formulation of source theory heavily relies on the energy-dependent Green function $G(\bi{r},\bi{r}';E)$ of the Hamiltonian. In the continuous spectrum of $H$, however, this integral kernel is by no means unique. From a physical point of view, this ambiguity of the solution set (\ref{eq:Wavefunction}) is required to accommodate different boundary conditions for the resulting wave function $\psi(\bi r)$. Here, we are interested in outgoing matter waves that enforce the use of the retarded energy Green function \cite{Economou83}. (This choice of kernel guarantees the correct sign for the total current $J(E)$ (\ref{eq:CurrentTotal}), as shown by an eigenfunction expansion of $G(\bi{r},\bi{r}';E)$ \cite{Bracher97}.) For reasons of causality, the retarded energy Green function is connected to the propagator $K(\bi{r},t|\bi{r}',0)$ of the equivalent time-dependent problem by the Laplace transform:
\begin{equation}\label{eq:GreenProp}
G(\bi{r},\bi{r}';E)=\frac{1}{\rmi\hbar} \int_0^\infty \rmd t\,\rme^{\rmi Et/\hbar}\,K(\bi{r},t|\bi{r}',0).
\end{equation}
The time-dependent quantum propagator is a thoroughly studied subject, and a fairly exhaustive list of available solutions is given in \cite{Kleber94,Grosche98}. In contrast, only few closed analytic solutions are known for energy-dependent Green functions in three dimensions. Notable examples are the expressions for the free field environment, the static uniform electric field \cite{Dalidchik76,Li90,Bracher98}, the static uniform magnetic field \cite{Bakhrakh71,Gountaroulis72,Dodonov75}, and combined parallel static electric and magnetic fields \cite{Fabrikant91,Kramer01}.

\subsection{A sum rule for the total current}

Exploiting the time-reversal symmetry relation for the propagator $K(\bi{r},t|\bi{r}',0)^\dagger=K(\bi{r},-t|\bi{r}',0)$, the total current $J(E)$~(\ref{eq:CurrentTotal}) can be rewritten using equation~(\ref{eq:GreenProp}) as
\begin{equation}\label{eq:CurrentTotalAlt}
J(E)=\frac{1}{\hbar^2}
\int_{-\infty}^{\infty}\rmd t\,\rme^{\rmi E t/\hbar}
\int\rmd^3\bi{r}
\int\rmd^3\bi{r}'
\,\sigma(\bi{r})^{*}\,K(\bi{r},t|\bi{r}',0)\,\sigma(\bi{r}').
\end{equation}
Integration with respect to the energy $E$, together with the initial condition $K(\bi{r},0|\bi{r}',0)=\delta(\bi{r}-\bi{r}')$, leads to the following sum rule for the total current:
\begin{eqnarray}\label{eq:CurrentSum}
\fl
\int_{-\infty}^{\infty}\rmd E\,J(E)
=\frac{2\pi}{\hbar}\int\rmd^3\bi{r}\int\rmd^3\bi{r}'
\,\sigma(\bi{r})^{*} K(\bi{r},0|\bi{r}',0)\,\sigma(\bi{r}')
=\frac{2\pi}{\hbar}\int\rmd^3\bi{r}\,{|\sigma(\bi{r})|}^2.
\end{eqnarray}

\section{Sources in a uniform force field}\label{sec:SourceUniformField}

To illustrate the formalism developed in the preceding section, we discuss the specific example of a quantum source of Gaussian shape embedded in a uniform force field $\bi{F}$. Experimentally, this situation is approximately realized in near-threshold photodetachment of negative ions in an electric field \cite{Bryant87,Baruch92,Blondel96,Blondel99,Gibson01}, where the limit of point-like sources applies. Considerably extended sources emerge when ultra cold Rubidium atoms are continuously released from a trapped Bose-Einstein condensate under the influence of gravity (``atom lasers'', \cite{Mewes97,Bloch99}).

\subsection{Propagator and the energy Green function}

The Hamiltonian $H_{\rm cont}$ for uniformly accelerated motion with $\bi{A}(\bi{r})=\bi{o}$ is given by:
\begin{equation}
H_{\rm field}=\frac{\bi{p}^2}{2m}-\bi{r}\cdot\bi{F}.
\end{equation}
For simplicity, we align the field along the $z$-axis: $\bi{F}=F\,{\hat \bi{e}}_z$. The propagator assigned to this Hamiltonian is well known and reads \cite{Kleber94,Grosche98,Kennard27}:
\begin{eqnarray}\label{eq:KField}
\fl
K_{\rm field}(\bi{r},t|\bi{r}',0)=
   {\left(\frac{m}{2\pi\rmi\hbar t}\right)}^{3/2}
   \exp\left\{ \frac{\rmi}{\hbar}\left[\frac{m}{2t}|\bi{r}-\bi{r}'|^2+\frac{Ft}{2}(z+z')-\frac{F^2 t^3}{24m}\right]\right\}.
\end{eqnarray}
The energy-dependent Green function is also available (see \cite{Dalidchik76,Li90,Bracher98} for a compact derivation) and may be expressed in terms of Airy functions \cite{Abramowitz65}:
\begin{equation}\label{eq:GreenField}
\fl
G_{\rm field}(\bi{r},\bi{r}';E) =\frac{m}{2\hbar^2}\frac{1}{|\bi{r}-\bi{r}'|}\left[{\rm Ci}(\alpha_+){\rm Ai}'(\alpha_-)-{\rm Ci}'(\alpha_+){\rm Ai}(\alpha_-)\right],
\end{equation}
where $\alpha_\pm=-\beta\left[2E+F(z+z')\pm F|\bi{r}-\bi{r}'|\right]$, $\beta=\left[m/(4\hbar^2F^2)\right]^{1/3}$, and ${\rm Ci}(x)={\rm Bi}(x)+\rmi\, {\rm Ai}(x)$. For further reference, we note that this Green function is invariant with respect to simultaneous shifts of the origin and the energy:
\begin{equation}\label{eq:GreenSymmetry}
G_{\rm field}(\bi{r},\bi{r}';E)=G_{\rm field}(\bi{r}-\bi{r}',\bi{o};E+Fz').
\end{equation}
For a compact notation it is suitable to introduce a set of scaled parameters
\begin{eqnarray}
\xi      =\beta F x                          \qquad &
\brho    =\beta F \bi{r}                     \nonumber\\
\nu      =\beta F y                          \qquad &
\epsilon =-2\beta E                          \\
\zeta    =\beta F z                          \qquad &
\tau     =t/(2\hbar\beta).                   \nonumber
\end{eqnarray}
This allows us to express the Green function in integral form via (\ref{eq:GreenProp}) and (\ref{eq:KField}):
\begin{equation}\label{eq:GreenTime}
\fl
G_{\rm field}(\brho,\bi{o};\epsilon)=-2\rmi\beta{(\beta F)}^3
\int_0^\infty \frac{\rmd\tau}{{(\rmi\pi\tau)}^{3/2}}\,
\exp\left(\frac{\rmi}{\tau}\rho^2+\rmi\tau(\zeta-\epsilon)-\frac{\rmi\tau^3}{12}\right).
\end{equation}

\subsection{Point-like sources}
\label{sec:DeltaSource}

Initially, we discuss a point-like quantum source in a homogeneous force field $\bi{F}$. Assuming an isotropic emission pattern, we may idealize the source $\sigma(\bi{r})$ in terms of the Dirac $\delta$-distribution
\begin{equation}\label{eq:SourceDelta}
\sigma_\delta(\bi{r})=C \delta(\bi{r}),
\end{equation}
where $C$ denotes some complex constant. We note that for this type of sources the sum rule stated in equation~(\ref{eq:CurrentSum}) is not applicable, since the ${\cal L}^2$ norm of $\delta(\bi{r})$ is not defined. Instead, the detailed mechanism behind the source in equation~(\ref{eq:SourceDelta}) merely influences the emission process through the scaling parameter $C$, the source strength. The wave function generated by $\sigma_\delta(\bi{r})$ directly follows from equations~(\ref{eq:Wavefunction}) and (\ref{eq:GreenField})
\begin{equation}
\psi(\bi{r})=C\,G_{\rm field}(\bi{r},\bi{o};E)
\end{equation}
and yields analytic expressions for the current density (\ref{eq:CurrentDensity}) in the field direction
\begin{equation}\label{eq:CurrentDensityPoint}
\fl
j_z(\bi{r},E)={|C|}^2\frac{m\beta F}{2\pi\hbar^3r^3}\left\{
z{{[\rm Ai}'(\alpha_-)]}^2+\beta\left[z(2E+Fz)+Fr^2\right] {[{\rm Ai}(\alpha_-)]}^2\right\},
\end{equation}
and for the total current (\ref{eq:CurrentTotal})
\begin{eqnarray}\label{eq:CurrentTotalPoint}
J(E)&=-\frac{2{|C|}^2}{\hbar}\lim_{r\rightarrow 0}{\rm Im}\left\{G_{\rm field}(\bi{r},\bi{o};E)\right\} \nonumber\\
    &=\frac{2{|C|}^2 m\beta F}{\hbar^3}\left\{
{[{\rm Ai}'(-2\beta E)]}^2+2\beta E {[{\rm Ai}(-2\beta E)]}^2\right\}.
\end{eqnarray}
In a different context, these expressions are implicitly contained also in \cite{Li90,Fabrikant91}.

\subsection{Gaussian source}
\label{sec:GaussianSource}

Now we turn our attention to more realistic spatially extended sources. Specifically, we consider a source term derived from the wave function $\psi_0(\bi{r})$ in equation (\ref{eq:Psi0}) of isotropic Gaussian shape
\begin{equation}\label{eq:SourceGauss}
\sigma(\bi{r})=\hbar\Omega\psi_0(\bi{r})=\hbar\Omega N_0 \exp(-r^2/(2a^2)).
\end{equation}
The parameter $a$ describes the width of the source and $N_0=a^{-3/2}\pi^{-3/4}$ denotes the proper normalization from the condition
\begin{equation}
\int\rmd^3\bi{r}\,{|\psi_0(\bi{r})|}^2=1.
\end{equation}
To obtain the expressions for the currents generated by a Gaussian source, we start with the derivation of the wave function from equation~(\ref{eq:Wavefunction}). Working in the time-dependent propagator representation (see equation~(\ref{eq:GreenProp})), we can carry out the $\bi{r}'$ integrations of Gaussian type in 
\begin{equation}
\psi(\bi{r})=-\rmi\Omega N_0\int_0^\infty \rmd t\,\rme^{\rmi E t/\hbar}\int \rmd^3\bi{r}'\,K_{\rm field}(\bi{r},t|\bi{r}',0)\,\rme^{-r'^2/(2a^2)}.
\end{equation}
Upon introducing the scaled width $\alpha=\beta F a$ and shifted parameters
\begin{equation}
\label{eq:ScaledParameter}
\tilde{\zeta}=\zeta+2\alpha^4 \qquad
\tilde{\epsilon}=\epsilon+4\alpha^4,
\end{equation}
the problem is reduced to a single complex integration
\begin{equation}\label{eq:PsiGaussTime}
\fl
\psi(\bi{r})=
-2\,\rmi\Lambda(\tilde{\epsilon})\beta{(\beta F)}^3
\int_{-2\rmi\alpha^2}^{\infty}\frac{\rmd u}{{(\rmi\,\pi u)}^{3/2}}
   \exp\left(
              \frac{\rmi}{u}\tilde{\rho}^2
	     +\rmi u(\tilde{\zeta}-\tilde{\epsilon})
	     -\frac{\rmi u^3}{12}
       \right).
\end{equation}
Here, $\Lambda(\tilde{\epsilon})=\hbar\Omega {(2\sqrt{\pi}a)}^{3/2} \rme^{2\alpha^2(\tilde{\epsilon}-4\alpha^4/3)}$, and $\tilde{\rho}^2=\xi^2+\nu^2+\tilde{\zeta}^2$. We moved the temporal integration into the complex plane by substituting $u=\tau-2\rmi\alpha^2$. The representation chosen in equation~(\ref{eq:PsiGaussTime}) emphasizes the close relationship of $\psi(\bi{r})$ to the Green function $G_{\rm field}(\brho,\bi{o};\epsilon)$ in the form~(\ref{eq:GreenTime}). To evaluate this integral analytically, we split the path of integration into two sections, one along the real $u$-axis, the other one along the imaginary $u$-axis: $\psi(\bi{r})=\psi_{{\rm near}}(\bi{r})+\psi_{{\rm far}}(\bi{r})$. The contribution due to
\begin{eqnarray}
\psi_{{\rm near}}(\bi{r})&=-2\,\rmi\Lambda(\tilde{\epsilon})\beta{(\beta F)}^3
\int_{-2\rmi\alpha^2}^{0}\frac{\rmd u}{{(\rmi\,\pi u)}^{3/2}}
   \rme^{
              \rmi\tilde{\rho}^2/u
	     +\rmi u(\tilde{\zeta}-\tilde{\epsilon})
	     -\rmi u^3/12}\nonumber\\
&\sim\Lambda(\tilde{\epsilon})\frac{2\beta{(\beta F)}^3}{\pi^{3/2}}\frac{\sqrt{2} \alpha}{\tilde{\rho}^2}\rme^{-\tilde{\rho}^2/(2\alpha^2)}
\end{eqnarray}
is a purely real term which drops off very quickly with increasing distance $\tilde{\rho}$ from the source region. The more interesting far-field contribution $\psi_{{\rm far}}(\bi{r})$ can be exactly evaluated using the integral representation (\ref{eq:GreenTime}):
\begin{eqnarray}
\psi_{{\rm far}}(\bi{r})&=
-2\,\rmi\Lambda(\tilde{\epsilon})\beta{(\beta F)}^3
\int_{0}^{\infty}\frac{\rmd u}{{(\rmi\,\pi u)}^{3/2}}
   \rme^{
              \rmi\tilde{\rho}^2/u
	     +\rmi u(\tilde{\zeta}-\tilde{\epsilon})
	     -\rmi u^3/12}\nonumber\\
&=\Lambda(\tilde{\epsilon})\,G_{\rm field}(\tilde{\brho},\bi{o};\tilde{\epsilon}).
\end{eqnarray}
With the help of equation~(\ref{eq:GreenSymmetry}) we can cast the last equation into the form
\begin{equation}
\label{eq:PsiVirtual}
\fl
\psi_{\rm far}(\bi{r})=\hbar\Omega {(2\sqrt{\pi}a)}^{3/2} \rme^{-m a^2 E/\hbar^2+m^2 F^2 a^6/(3\hbar^4)}
\,G_{\rm field}(\bi{r},-\frac{m\bi{F}}{2\hbar^2}a^4;E).
\end{equation}
\begin{figure}[t]
\begin{center}
\includegraphics[width=\textwidth]{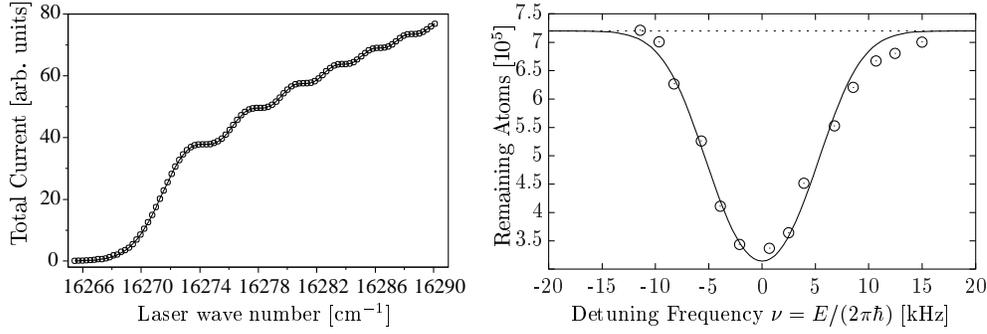}
\end{center}
\caption{\label{fig:jtot_exp}
Characteristics of matter waves from sources in uniform force fields.\\
Left panel: Total current $J(E)$ in near-threshold photodetachment of S$^-$ in a homogeneous electric field $F=2.205\cdot10^4\,$eV/m. (\opencircle) Experimental data of Gibson \etal \cite{Gibson01}. (\full) Theoretical result from (\ref{eq:CurrentTotalPoint}) with adjusted source strength $C$. Right panel: Number of atoms remaining in a Bose-Einstein condensate $N(T)$ after continuous release of atoms for $T=20\,$ms as a function of the detuning frequency $\nu$. (\opencircle) Measurement by Bloch \etal reported in \cite{Gerbier01}. (\full) Theoretical prediction according to (\ref{eq:CurrentTotalGauss}) and (\ref{eq:RemainingAtoms}) with effective Gaussian condensate width $a=2.8\,\mu$m and outcoupling strength $\Omega=2\pi\cdot105.585\,$Hz.}
\end{figure}
This expression displays a remarkable feature of the wave function $\psi_{\rm far}(\bi{r})$ originating from a Gaussian source: The extended Gaussian source can be replaced by a virtual point source of the same energy at a location shifted by $m\bi{F}a^4/(2\hbar^2)$ from the centre of the Gaussian distribution, carrying the energy dependent weight $\Lambda(\tilde{\epsilon})$. From this relation, the expression for the currents due to (\ref{eq:SourceGauss}) are conveniently found from the analogous expressions for a point source by just performing the indicated shifts. Neglecting $\psi_{\rm near}(\bi{r})$, the far-field current density reads according to equation~(\ref{eq:CurrentDensityPoint})
\begin{eqnarray}\label{eq:CurrentDensityGauss}
\fl
j_z(\brho,\tilde{\epsilon})=
16\sqrt{\pi}\hbar\Omega^2\beta^3 F^2 
\rme^{4\alpha^2(\tilde{\epsilon}-4\alpha^4/3)}\nonumber\\
\lo\times\frac{\alpha^3}{\tilde{\rho}^3}
\left\{
\tilde{\zeta}
{[{\rm Ai}'(\tilde{\epsilon}-\tilde{\zeta}+\tilde{\rho})]}^2
+\beta\left[\tilde{\zeta}(\tilde{\zeta}-\tilde{\epsilon})+\tilde{\rho}^2\right]
{[{\rm Ai} (\tilde{\epsilon}-\tilde{\zeta}+\tilde{\rho})]}^2\right\}.
\end{eqnarray}
The same procedure yields the total current. However, since both $\psi_{\rm near}(\bi{r})$ and $\sigma(\bi{r})$ are purely real and only the imaginary part of $\psi(\bi{r})$ is needed for the evaluation of the total current (see equation~(\ref{eq:CurrentDivergence})), the following expression obtained by shifting the energy in equation~(\ref{eq:CurrentTotalPoint}) is even an exact result:
\begin{equation}
J(\tilde{\epsilon})\label{eq:CurrentTotalGauss}
=64\pi^{3/2}\hbar\Omega^2\alpha^3\beta
\rme^{4\alpha^2(\tilde{\epsilon}-4\alpha^4/3)}\left\{
{[{\rm Ai}'(\tilde{\epsilon})]}^2-\tilde{\epsilon} {[{\rm Ai}(\tilde{\epsilon})]}^2\right\} .
\end{equation}
In the limit of extended Gaussian sources, a simple approximation to this formula is available that leads to a geometrical interpretation. We start out with a time dependent integral formulation of the total current that follows from equations~(\ref{eq:CurrentTotal}) and (\ref{eq:GreenTime}) after the spatial integrations are performed:
\begin{equation}
J(\tilde{\epsilon})
=32\hbar\Omega^2\alpha^3\beta
\rme^{4\alpha^2(\tilde{\epsilon}-4\alpha^4/3)}
{\rm Im}\left\{
\int_{0}^{\infty}\frac{\rmi\, \rmd u}{{(\rmi u)}^{3/2}}
\rme^{-\rmi u \tilde{\epsilon}-\rmi u^3/12} \right\}.
\end{equation}
The integral is evaluated in saddle point approximation. Assuming $\alpha\gg\epsilon$, we keep only the leading order terms of a Taylor expansion in the exponent and prefactor. The resulting current has Gaussian shape:
\begin{equation}
J_{\rm sp}(\epsilon)
=\frac{2\sqrt{\pi}\hbar\Omega^2\beta}{\alpha}\rme^{-\epsilon^2/(4\alpha^2)}.
\end{equation}
This expression is equivalent to the implicit representation
\begin{equation}\label{eq:CurrentTotalSP}
J_{\rm sp}(E)
=\frac{2\pi}{\hbar}\int\rmd^3\bi{r}\,{|\sigma(\bi{r})|}^2\,\delta(E+F z).
\end{equation}
Evidently, the approximation (\ref{eq:CurrentTotalSP}) obeys the sum rule (\ref{eq:CurrentSum}) for the total current $J(E)$. For extended sources, the energy dependence of $J_{\rm sp}(E)$ reflects the source structure: By the resonance condition $E+F z=0$, the total current probes the density ${|\psi_0(\bi{r})|}^2$ on different slices across the source.
\begin{figure}[t]
\begin{center}
\includegraphics[width=\textwidth]{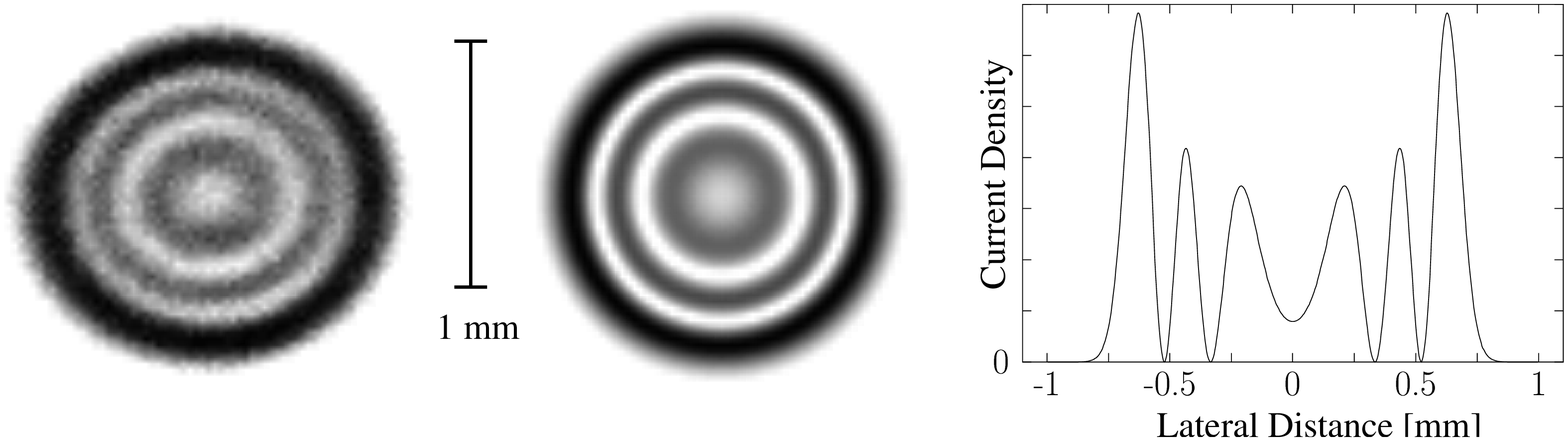}
\end{center}
\caption{\label{fig:jz_photo}
Interference fringes in uniformly accelerated motion.\\
Distribution of electrons in near-threshold photodetachment of O$^-$ in a homogeneous electric field $F=423\,$eV/m. Distance from the source $z=0.514\,$m, electron energy $E=100.5\,\mu$eV. Left panel: Image recorded by Blondel \etal \cite{Blondel99}. Centre panel: Calculation from (\ref{eq:CurrentDensityPoint}). Right panel: Corresponding current density profile as a function of the lateral distance.}
\end{figure}

\section{Examples: Photodetachment and atom laser}\label{sec:Examples}

As an application for the developed formalism, we want to discuss two physical systems that have recently undergone experimental evaluation.

\subsection{Photodetachment}

In photodetachment experiments, negatively charged ions are illuminated by a laser beam to release the surplus electron.  If the photon energy closely matches the electron affinity of the ionic species, the detached electron starts out with a well-defined minuscule amount of kinetic energy $E$.  Then, the size of the emitting ion is small compared to the initial electronic wavelength $\lambda=h/\sqrt{2mE}$, and details of the atomic structure cannot be resolved:  The detached electron wave is solely characterized by its orbital angular momentum (which is in turn fixed by the selection rules of the underlying dipole transition), and allows for a description in terms of a point source $\sigma(\bi r)$.  For isotropic ($s$--wave) emission, the Dirac $\delta$--distribution $\sigma_\delta(\bi r)$ (\ref{eq:SourceDelta}) is a viable choice.

In a field-free environment, the electron is emitted simply into a spherical wave, $\psi(\bi r)\propto \rme^{\rmi kr}/r$.  In this case, the quantum source model presented in section~\ref{sec:SourceModel} recovers the well-established behaviour of the photodetachment cross section near threshold known as Wigner's law \cite{Wigner48} which states that $J(E)\propto \sqrt{E}$. Recently, the question how the threshold behaviour of the photocurrent is altered in the presence of external fields in the interaction region has found considerable attention \cite{Bryant87,Larson78}, and precision experiments were conducted to measure both the energy dependence of the cross section \cite{Baruch92,Gibson01} and the corresponding photoelectron distribution \cite{Blondel96,Blondel99} for photodetachment in a homogeneous electric field environment.  Here, the electric force ${\bi F}=-e{\bi E}$ accelerating the detached electrons allows for direct comparison with the results presented in section~\ref{sec:DeltaSource}.  The left panel in figure~\ref{fig:jtot_exp} displays the photocurrent measured by Gibson \etal \cite{Gibson01} together with the source-theoretical prediction (\ref{eq:CurrentTotalPoint}) first obtained in a different fashion by Fabrikant \cite{Fabrikant91}.  The agreement is striking.  Besides a slow onset below threshold indicating tunneling emission, the plot prominently features a ``staircase'' appearance of $J(E)$ modifying Wigner's law.  Although an explanation of this electric field induced effect in terms of the closed free-falling orbit has been offered \cite{DuDelos89}, it is more accurately described as the remaining imprint of a remarkable interference pattern in the current density distribution $j_z({\bi r},E)$ (\ref{eq:CurrentDensityPoint}).

In a recent series of experiments \cite{Blondel96,Blondel99}, Blondel \etal recorded the spatial distribution of the photocurrent on a distant detector plane perpendicular to the applied electric field. One of the impressive images obtained using the ``photodetachment microscope'' is depicted in figure~\ref{fig:jz_photo} (left panel), revealing an interference ring pattern of macroscopic size. In the source formalism, the electronic distribution should reflect the local current density $j_z({\bi r},E)$ pertaining to the Green function $G_{\rm field}(\bi{r},\bi{o},E)$ (\ref{eq:GreenField}) in a uniform force field. Indeed, an image calculated from (\ref{eq:CurrentDensityPoint}) closely resembles the experimental result (centre panel in figure~\ref{fig:jz_photo}). For comparison also the corresponding lateral current profile is shown (right panel). 

The conspicuous fringes find a simple semiclassical interpretation in terms of two-path interference: As first pointed out by Demkov \cite{Demkov81}, within the sector of classically allowed motion always two ballistic trajectories will connect the origin with a given location on the detector. Hence, the uniform electric field effectively acts as a two-slit interferometer \cite{Bracher98}, providing a sensitive device for the precise determination of electron affinities \cite{Blondel99}. Similar results are expected for point sources in parallel electric and magnetic fields where four-path interference takes place \cite{Kramer01}.
\begin{figure}[t]
\begin{center}
\includegraphics[width=\textwidth]{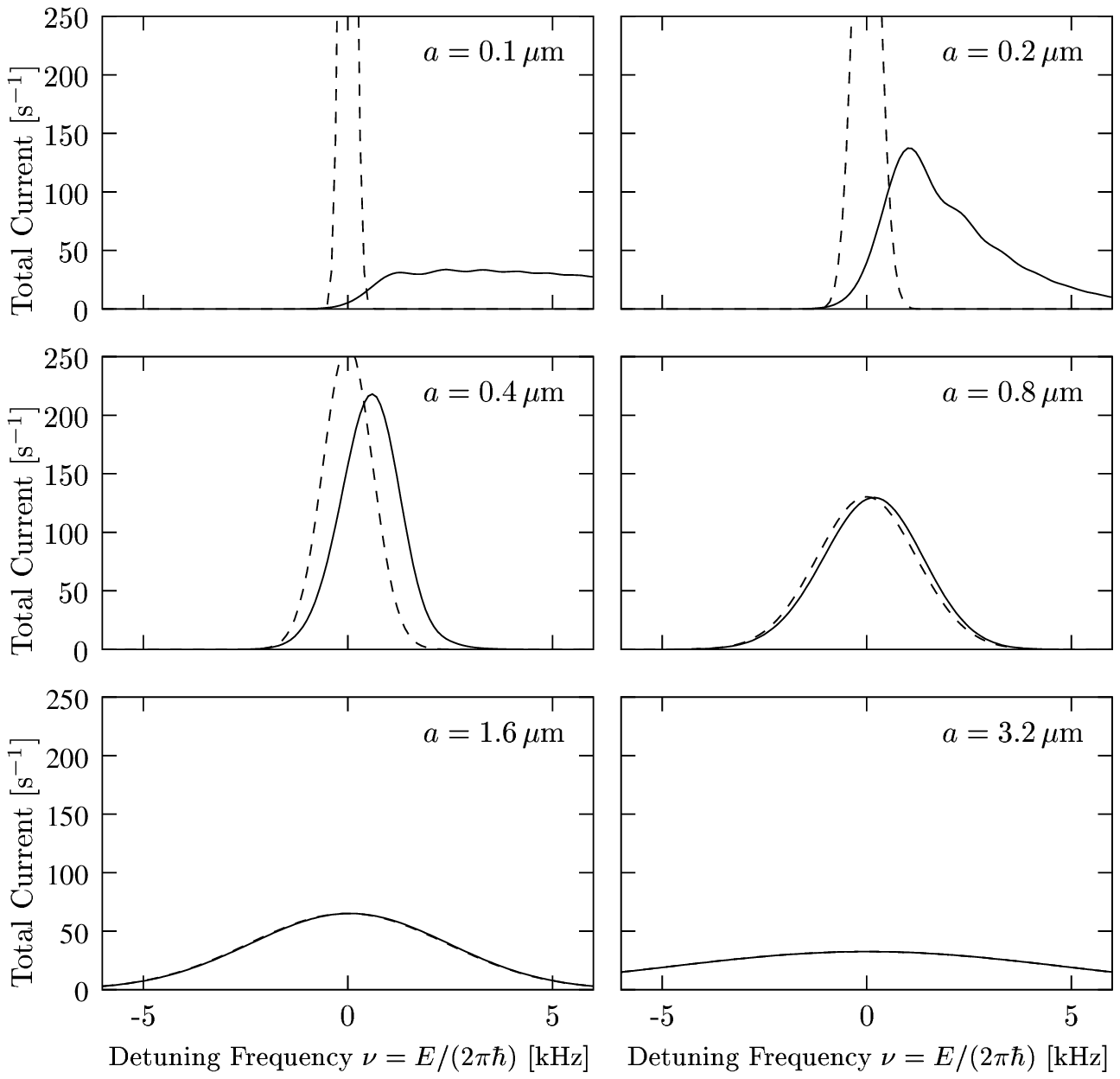}
\end{center}
\caption{\label{fig:jtot_gauss}
Transition from a point-like source to an extended source.\\
Depicted is the total current $J(E)$ from a Gaussian source of freely falling Rb atoms for different source widths $a$ versus the detuning frequency $\nu$. (\full) Exact currents from (\ref{eq:CurrentTotalGauss}). (\dashed) Slicing approximation according to (\ref{eq:CurrentTotalSP}). The coupling strength is $\Omega=2\pi\cdot 100\,$Hz. Beam profiles are shown in figure~\ref{fig:jz_gauss}.}
\end{figure}
\begin{figure}[t]
\begin{center}
\includegraphics[width=\textwidth]{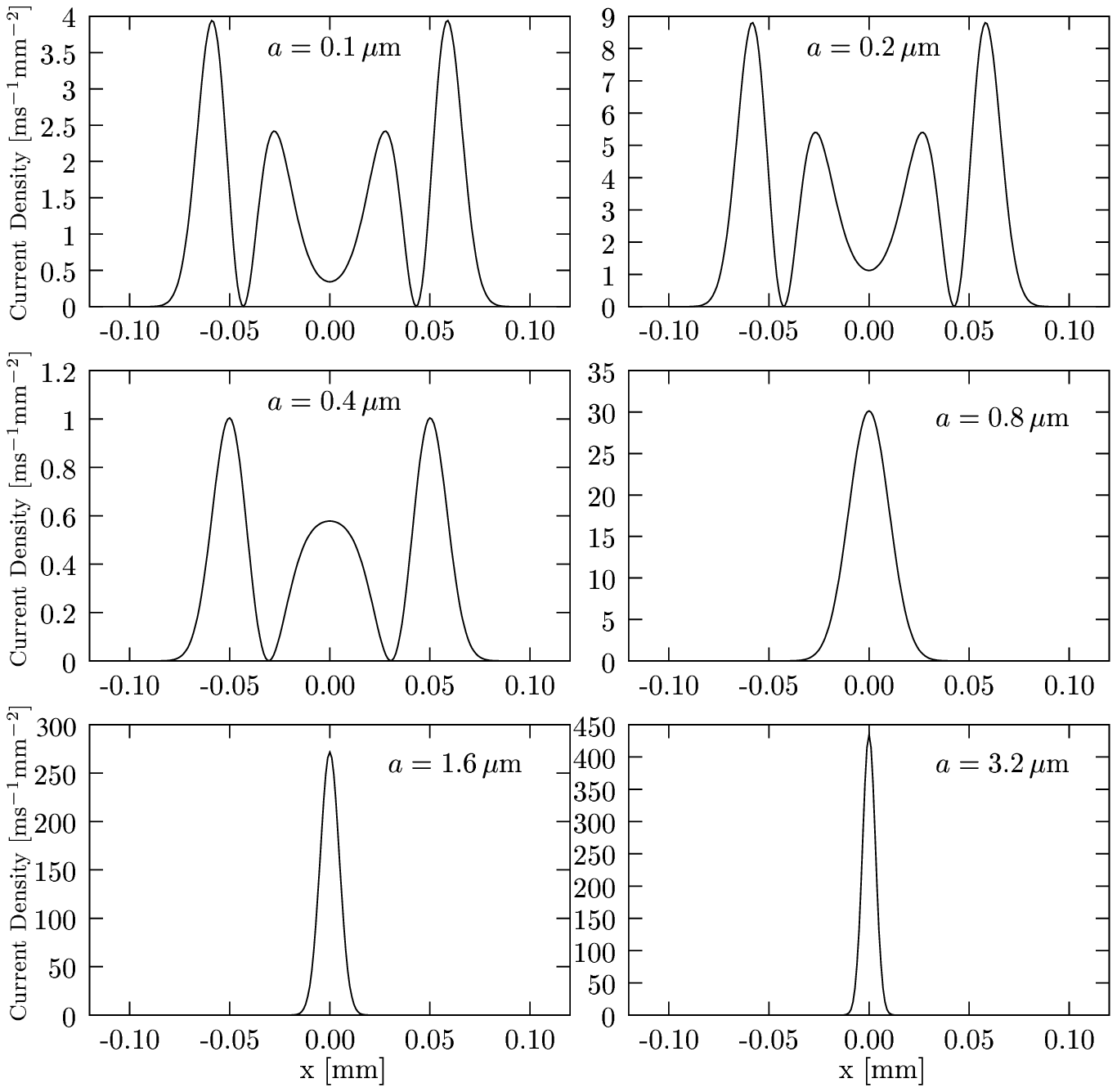}
\end{center}
\caption{\label{fig:jz_gauss}
Size dependence of the particle distribution.\\
Profiles of the current density $j_z(x,0,z,E=2\pi\nu\hbar)$ for Gaussian sources emitting free-falling Rb atoms. All results are given for $\nu=2.5\,$kHz and $\Omega=2\pi\cdot 100\,$Hz. There is rotational symmetry about the $z$-axis (see figure~\ref{fig:jz_photo}). The profiles are calculated from (\ref{eq:CurrentDensityGauss}) at a distance $z=1\,$mm from the source. The source width is denoted by $a$. See also figure~\ref{fig:jtot_gauss}.}
\end{figure}

\subsection{Atom laser}

In section~\ref{sec:SourceModel} we introduced quantum sources using a coupled two state model. This model may serve as a simplified description of the atom laser, a coherent beam of particles released from a trapped Bose-Einstein condensate (BEC) \cite{Mewes97,Bloch99}. The two states are realized by different hyperfine Zeeman levels of Rb. The state $\psi_{\rm trap}(\bi{r})$ is trapped by a magnetic field, whereas the continuum state $\psi_{\rm cont}(\bi{r})$ is not influenced by the trap but subject to gravitational attraction. These states are coupled through an oscillating magnetic field of adjustable frequency with a specific coupling strength $\hbar\Omega$. Near its minimum, the potential for the trapped state is approximately harmonic. For non-interacting particles, the ground state of the condensate in this potential is given by a oscillator wave function. In our discussion, we maintain the Gaussian shape of the condensate wave function $\psi_0(\bi{r})$, but increase its width $a$ in order to account for the repulsive atomic interaction.

The ensuing quantum source $\sigma(\bi{r})$ continuously releases Rb atoms (mass $m$) with an initial energy $E=2\pi\hbar\nu$ determined by the frequency of the perturbation. These particles are subsequently accelerated by the gravitational  force $F=mg$ ($g\approx9.81\,$m/s$^2$). From section~\ref{sec:GaussianSource}  it is possible to predict the efficiency of this process (the total particle current) as well as the spatial distribution of atoms (their current density profile). Experimental data concerning the number of remaining condensate atoms $N(T)$ after $T=20\,$ms of atom laser operation is available \cite{Gerbier01}. Obviously, this number is related to the current by
\begin{equation}\label{eq:RemainingAtoms}
N(T)=N(0)\exp[-J(E) T],
\end{equation}
where $N(0)$ denotes the initial number of atoms in the BEC and $J(E)$ the total current. Besides the gravitational force, the relevant parameters entering the theoretical prediction for $J(E)$ in equation~(\ref{eq:CurrentTotalGauss}) are the coupling strength $\hbar\Omega$ and the Gaussian width $a$. In figure~\ref{fig:jtot_exp} (right panel), the calculated number of remaining atoms (\ref{eq:RemainingAtoms}) is compared to the experimental measurement by Bloch \etal reported in \cite{Gerbier01}. The coupling frequency $\Omega$ used for the calculation  is fixed by the sum rule (\ref{eq:CurrentSum}) applied to the experimental data, and the effective width $a=2.8\,\mu$m (that is actually largely governed by atomic repulsion in the BEC) is obtained from a fit. Contrary to the case of photodetachment, the current characteristics faithfully reproduces the Gaussian shape of the source, as stated by the approximation obtained by slicing the condensate at height $z=E/F$ (\ref{eq:CurrentTotalSP}). However, according to the exact expression for $J(E)$ (\ref{eq:CurrentTotalGauss}), source theory predicts a dramatic change in behaviour of the total current for smaller condensate sizes as illustrated in figure~\ref{fig:jtot_gauss}. The plots show both the analytic solution $J(E)$ and the corresponding approximation $J_{\rm sp}(E)$ for various source widths $a$. According to the sum rule (\ref{eq:CurrentSum}), the area under all curves is the same. For $a>1\,\mu$m, both models yield almost identical curves. However, for smaller condensates the differences become noticeable and for $a<0.4\,\mu$m, the slicing model fails completely: $J(E)$ becomes asymmetric with respect to $\nu=0$, and oscillations in the current are prominent. These features are familiar from the photocurrent discussed in the previous section, where we related them to two-path interference in a uniform field (figure~\ref{fig:jz_photo}). 

To examine if similar interference fringes occur also in an ideal atom laser beam, we compute the current density profiles for the set of Gaussian sources displayed in figure~\ref{fig:jtot_gauss} from (\ref{eq:CurrentDensityGauss}). The results from this calculation are shown in figure~\ref{fig:jz_gauss} where we choose as initial energy $E=2\pi\hbar\nu$ with $\nu=2.5\,$kHz. The distance from the source is fixed at $z=1\,$mm. In this figure, a distinct ring pattern in the current density prevails for $a\le 0.4\,\mu$m. The number of fringes diminishes with increasing source width, until for $a\ge 0.8\,\mu$m the current profile attains an increasingly narrow Gaussian shape. 
To interpret this behaviour, we first note that for an extended source region, the simple concept of two interfering paths originating from a single point in space is not readily applicable. Recalling the particular property of a Gaussian source to act as a virtual point source shifted in space (\ref{eq:PsiVirtual}), we may recover the concept of interfering paths. However, the effective initial kinetic energy decreases with growing source size (\ref{eq:ScaledParameter}) and becomes negative for $E<mF^2a^4/(2\hbar^2)$, leading to a ``virtual'' tunneling source that emits a beam of Gaussian profile. Its properties are discussed in detail in \cite{Bracher98}.

\section{Conclusion and outlook}\label{sec:Outlook}

Quantum sources that are inserted into the Schr\"odinger equation provide a convenient and practical tool to assess and solve scattering problems in external potentials.  Specifically, we studied point-like and Gaussian quantum sources embedded within a three-dimensional homogeneous force field, and analytical solutions for the resulting particle distribution and total current were obtained.  For point sources, the model is in excellent agreement with data gained from near-threshold photodetachment experiments conducted in the presence of an electric field. Available measurements on continuous atom laser beams are in reasonable accordance with the theory presented for an extended source. Dependent on the condensate size, we predict the appearance of a ring pattern related to two-path interference in the atom laser beam released from a single ideal Bose-Einstein condensate.

A modification of the presented theory to comprise sources with definite angular momentum is feasible and will be the subject of a forthcoming publication.
Physical applications of this extension include photodetachment experiments involving $p$-wave emission and the effects of vortices in an ideal Bose-Einstein condensate on the outcoupled atom laser beam.

\ack{
Experimental data on near-threshold photodetachment experiments was kindly provided by C.~Blondel, N.~D.~Gibson and C.~W.~Walter. We appreciate useful discussions with these authors as well as P.~Kramer, W.~Becker, T.~Esslinger and T.~H\"ansch. We also thank Kilian Bracher for his patience during the preparation of this paper. Partial financial support by the Deutsche Forschungsgemeinschaft is gratefully acknowledged.}

\end{document}